\documentclass[10pt,conference]{IEEEtran}
\IEEEoverridecommandlockouts
\usepackage{cite}
\usepackage{amsmath,amssymb,amsfonts}
\usepackage{algorithmic}
\usepackage{graphicx}
\usepackage{textcomp}
\usepackage{xcolor}
\usepackage{multirow}
\usepackage[table]{xcolor}
\usepackage{tcolorbox}
\usepackage{array}
\usepackage{caption}
\usepackage{url}
\usepackage[switch]{lineno}
\usepackage[acronym]{glossaries}
\usepackage{microtype}
\makeglossaries

\newacronym{cti}{CTI}{Cyber Threat Intelligence}
\newacronym{llm}{LLM}{Large Language Model}
\newacronym{llms}{LLMs}{Large Language Models}
\newacronym{nl}{NL}{Natural Language}
\newacronym{stix}{STIX}{Structured Threat Information Expression}
\newacronym{cav}{CAV}{Connected and Autonomous Vehicles}
\newacronym{cve}{CVE}{Common Vulnerabilities and Exposures}
\newacronym{cwe}{CWE}{Common Weakness Enumeration}
\newacronym{nvd}{NVD}{National Vulnerability Database}
\newacronym{ttps}{TTPs}{Tactics, Techniques, and Procedures}
\newacronym{sdo}{SDO}{STIX Domain Objects}
\newacronym{sro}{SRO}{STIX Relationship Objects}
\newacronym{nlp}{NLP}{Natural Language Processing}
\newacronym{ev}{EV}{Electric Vehicle}

\usepackage[numbers,sort&compress]{natbib} 
\newcommand{\mycomment}[2]{#2}

\def\BibTeX{{\rm B\kern-.05em{\sc i\kern-.025em b}\kern-.08em
    T\kern-.1667em\lower.7ex\hbox{E}\kern-.125emX}}
\begin{document}
\mycomment{\title{Benchmarking Open-Weight LLM for STIX Generation from Connected and Autonomous Vehicle-Related Vulnerabilities}}{\title{\title{\title{Evaluating Open-Weight LLMs for Generating Structured Threat Information for Autonomous Vehicle Vulnerabilities}}}
}

\author{Md Erfan, Ahmed Ryan, Md Kamal Hossain Chowdhury and Md Rayhanur Rahman\textsuperscript{$\dagger$}%
\thanks{Md Erfan, Ahmed Ryan, and Md Rayhanur Rahman are with the Department of Computer Science, The University of Alabama, Tuscaloosa, USA. Email: \{merfan, aryan9\}@crimson.ua.edu, mrahman87@ua.edu}%
\thanks{Md Kamal Hossain Chowdhury is with the Alabama Water Institute, The University of Alabama, Tuscaloosa, USA. Email: mhchowdhury@crimson.ua.edu}%
\thanks{\textsuperscript{$\dagger$}Corresponding author.}%
}
\maketitle

\begin{abstract}
Connected and Autonomous Vehicles (CAVs) rely on interconnected software and hardware components, including sensors, Electronic Control Units, in-vehicle infotainment systems, and telematics units, where vulnerabilities can compromise assets, users, and vehicle operations. These vulnerabilities are commonly documented as plain text in the Common Vulnerabilities and Exposures (CVE) database; however, security practitioners require structured information about affected assets, types of weaknesses, and attack behaviors to effectively mitigate the risks from these vulnerabilities. To this end, we evaluate open-weight Large Language Models (LLMs) for generating Structured Threat Information Expression (STIX), a well-known structured format for representing threat information, for CAV-related CVEs. We construct a dataset called \textit{CAV-STIXGen} that maps CAV vulnerability descriptions to STIX domain objects (SDO), STIX relationship objects (SRO), Common Weakness Enumeration (CWE), and MITRE ATT\&CK techniques mappings. Using this dataset, we evaluated 11 open-weight LLMs (4B to 120B parameters) across various prompting strategies and temperatures. Single-model configurations achieve F1 scores of 0.94 for SDO, 0.63 for SRO, and 0.99 for CWE mapping, while complete MITRE ATT\&CK mapping remains challenging. In a multi-agent setup, Gemma-4-31B and Codestral-22B achieve F1 scores of 0.91 for SDOs and 0.43 for SROs, respectively. Lastly, we analyze CWE and MITRE ATT\&CK co-occurrences to identify recurring threat patterns in the CAV domain, demonstrating how AI-assisted vulnerability-to-STIX translation can automate threat intelligence and prioritize defense in transportation security.

\end{abstract}

\begin{IEEEkeywords}
STIX Domain Objects, STIX Relationship Objects, CWE, Mitre Attack Techniques 
\end{IEEEkeywords}

\section{Introduction}
\gls{cav} operate through a complex interplay of software and hardware modules, such as sensors (e.g., LiDAR, radar, cameras), sensor fusion modules, Electronic Control Units, the internal Controller Area Network, in-vehicle infotainment systems, telematics control units, and Vehicle-to-Everything communication systems. All complex software systems contain bugs, and the extensive codebase of CAVs is no exception; a fraction of these bugs can be exploited as security vulnerabilities. Recent automotive security reports show that such vulnerabilities continue to affect connected vehicle services, with 2024 Kia vulnerabilities enabling remote vehicle control using only a license plate and 2025 Subaru STARLINK flaws exposing vehicle controls, customer data, and historical location records~\cite{curry2024kia,curry2025subaru}.

\begin{figure}[t]
    \centering
    \includegraphics[width=\columnwidth]{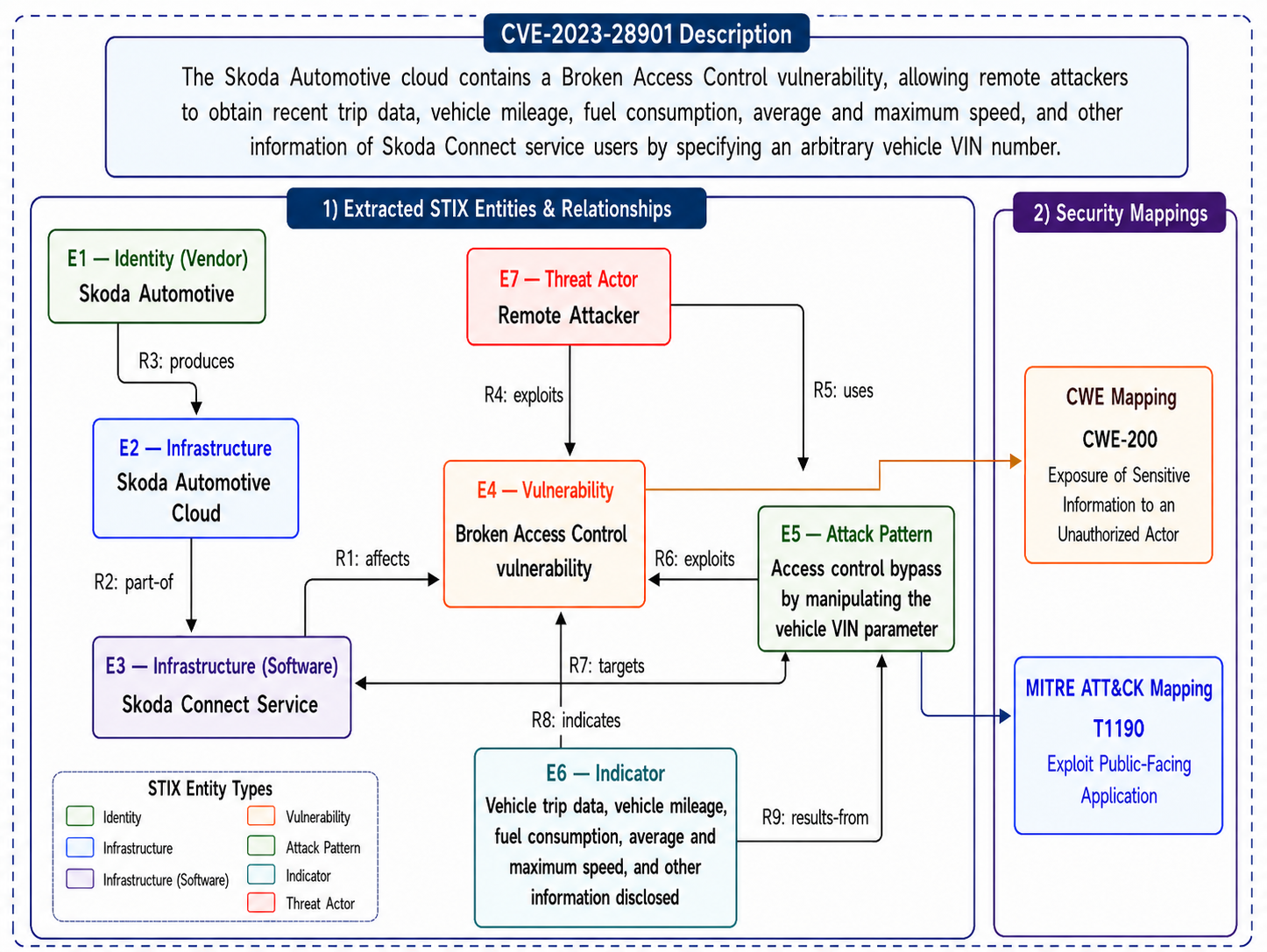}
    \caption{An Example of CVE Description to STIX Generation.}
    \label{fig:motivational_example}
\end{figure}

The information about these vulnerabilities is provided as plain text, which we refer to as \gls{cve} descriptions. As a result, security practitioners must analyze unstructured CVE descriptions to identify affected assets, weakness types, attack behaviors, and defensive actions. Converting CVE descriptions into a structured knowledge representation, such as Structured Threat Information Expression (STIX)~\cite{stix_project}, can organize vulnerability knowledge into entities, relationships, weakness mappings, and attack-behavior mappings, as shown in Fig.~\ref{fig:motivational_example}. Such structured representation helps practitioners compare vulnerabilities, analyze attack paths, and prioritize defenses more systematically.

\gls{llms} provide language-understanding and structured knowledge-generation capabilities, which make them relevant for CVE-to-STIX generation. However, prior studies on STIX generation from cybersecurity-related text show that LLMs-generated outputs can include incomplete objects, incorrect relationships, invalid structures, and hallucinated attribute values~\cite{papoutsis2025cti,siracusano2023time}. Therefore, a ground-truth dataset is necessary before LLM-generated STIX can be reliably used for security analysis. Evaluating LLMs requires a ground-truth dataset that connects CVE descriptions with STIX entities, relationships, \gls{cwe}, and MITRE ATT\&CK mappings. Existing datasets for STIX generation~\cite{marchiori2023stixnet,alam2024ctibench,ghosh2025cve} focus on generating MITRE ATT\&CK techniques and CWE identification from cybersecurity-related text. However, to the best of our knowledge, no CAV-focused dataset exists that maps CVE descriptions to STIX bundle generation.

\textit{The goal of this research is to help security practitioners conduct transportation-domain threat analysis by evaluating LLMs for CVE-to-STIX generation in \gls{cav}.}

We address the following research questions (RQ).
\begin{itemize}
    \item \textbf{RQ1 [Dataset Construction]:} How can we construct a dataset of STIX representations from unstructured \gls{cve} descriptions in the \gls{cav} domain?

    \item \textbf{RQ2 [Open-Weight LLM Evaluation]:} How effectively do open-weight \gls{llms} perform in CVE-to-STIX generation under different prompting strategies?

    \item \textbf{RQ3 [CWE and MITRE ATT\&CK Technique Analysis]:} What are the frequent \gls{cwe} and MITRE ATT\&CK techniques in CAV-related vulnerabilities, and which ATT\&CK techniques co-occur in the dataset?    
\end{itemize}

We construct a CAV-focused dataset by manually annotating STIX entities, relationships, CWE mappings, and MITRE ATT\&CK technique mappings, and then converting the validated annotations into ground-truth STIX JSON files. Using this dataset, we evaluate open-weight LLMs under multiple prompting strategies and measure performance across \gls{sdo} extraction, \gls{sro} extraction, CWE, and MITRE ATT\&CK mapping. We also analyze ATT\&CK technique co-occurrence in the ground-truth data to identify recurring attack-behavior patterns. 

We make the following contributions:

\begin{itemize}
    \item We construct a CAV-focused benchmark dataset by manually annotating CAV-related CVEs and converting the validated annotations into ground-truth STIX JSON files.

    \item We evaluate open-weight LLMs for generating structured STIX representations from unstructured CVE descriptions.

    \item We compare prompting strategies to examine how different levels of task guidance affect generation quality.

    \item We analyze MITRE ATT\&CK techniques in the ground-truth data to identify frequent techniques and recurring co-occurrence patterns in CAV-related vulnerabilities.
\end{itemize}

We make the dataset, prompts, and replication scripts available to independent researchers and practitioners (https://figshare.com/s/29f4ceff6bb1c6de5700).

The remainder of this paper is organized as follows. Section~\ref{subsec:background_key_concepts} introduces the key concepts used in this study, including CVE, CWE, MITRE ATT\&CK, and STIX. Section~\ref{sec:related_work} reviews prior work on cyber threat intelligence datasets, LLM-based STIX generation, and CWE and MITRE ATT\&CK technique analysis. Section~\ref{sec:methodology} describes the construction of the \textit{CAV-STIXGen} dataset, the LLM evaluation design, the CWE and MITRE ATT\&CK analysis, and the experimental setup. Section~V presents the findings for the research questions, including dataset characteristics, open-weight LLM performance, and recurring weakness and attack-behavior patterns in CAV-related vulnerabilities. Finally, Section~\ref{sec:conclusion} concludes the paper and discusses future research directions.

\section{Key Concepts}
\label{subsec:background_key_concepts}
In this section, we discuss key concepts related to this study.

\textbf{Common Vulnerabilities and Exposures (CVE)} is a public database for known vulnerabilities. Each \gls{cve} record provides a unique identifier, such as CVE-2023-28901\footnote{\url{https://nvd.nist.gov/vuln/detail/CVE-2023-28901}}, and a natural-language description of the vulnerability. CVE records can be found through the official CVE website\footnote{\url{https://www.cve.org/}} and through vulnerability databases such as the NVD\footnotemark[1]. \mycomment{CVE records can be found through the CVE website.}{}

\textbf{Common Weakness Enumeration (CWE)} is a standardized classification of software and hardware weakness types. \gls{cwe} mappings help identify the root weakness behind a vulnerability, such as improper access control, buffer overflow, injection, or improper input validation. 

\textbf{MITRE ATT\&CK} is a taxonomy for adversary behavior into \gls{ttps}. Tactics describe the attacker’s objective, such as initial access, execution, persistence, privilege escalation, or impact. Techniques describe the methods used to achieve those objectives, such as exploiting a public-facing application or causing endpoint denial of service. MITRE ATT\&CK helps security practitioners describe attacks in a consistent vocabulary and connect vulnerability information with known adversary behaviors.

\textbf{\gls{stix}} is a standardized format for representing and sharing cyberthreat information. Within STIX, cyber threat knowledge is organized through \gls{sdo} and \gls{sro}. We present a motivational example of converting CVE-2023-28901\footnotemark[1] into STIX in Fig.~\ref{fig:motivational_example}. The CVE description states that a broken access-control vulnerability in the Skoda Automotive cloud allows remote attackers to access Skoda Connect user information through an arbitrary vehicle Vehicle Identification Number (VIN). The STIX representation separates the plain-English description into entities, including the vendor, cloud infrastructure, affected service, vulnerability, attack pattern, indicator, and remote attacker. The relationships connect these entities by showing that the vulnerability affects the Skoda Connect service, the remote attacker exploits the vulnerability, and the attack pattern targets the affected service. The example also adds CWE-200 for sensitive information exposure and MITRE ATT\&CK technique T1190 for exploiting a public-facing application.\mycomment{This structured representation helps security practitioners identify the affected asset, weakness type, attack behavior, and exposed information more efficiently. STIX is formatted in JSON format.} {This structured representation helps security practitioners identify the affected asset, weakness type, attack behavior, and exposed information more efficiently. In STIX, the \gls{sdo} and \gls{sro} are packaged as a JSON-based STIX bundle for machine-readable storage, sharing, and analysis.}

\section{Related Work}
\label{sec:related_work}

We organize related work according to the main research directions of our study: dataset construction for CTI and vulnerability understanding, LLM-based STIX and CTI generation, and CWE and MITRE ATT\&CK pattern analysis.

\textbf{Dataset Construction for CTI and Vulnerability Understanding.}
Prior work has developed various datasets and benchmarks for cyber threat intelligence, vulnerability understanding, and security knowledge extraction. For instance, CTIBench~\cite{alam2024ctibench} provides a benchmark for evaluating LLMs on CTI tasks, including CWE prediction, MITRE ATT\&CK technique extraction, threat actor attribution, and CTI concept understanding. Focusing on vulnerabilities, CVE-LLM~\cite{ghosh2025cve} uses an ontology-assisted LLM approach for asset and vulnerability based evaluations, while Text2Weak~\cite{simonetto2024text2weak} maps CVE descriptions to CWE categories using description embeddings and LLMs. From a structural perspective, STIXnet~\cite{marchiori2023stixnet} constructs a pipeline for extracting STIX objects from CTI reports using rule-based methods, knowledge bases, Natural Language Processing (NLP), machine learning, and deep learning. Similarly, CTI-GEN~\cite{papoutsis2025cti} generates STIX 2.1-compliant CTI from textual reports and manually constructed attack scenarios. Notably, A recent survey on \gls{ttps} extraction reports that vulnerability databases are used less frequently than benchmark datasets, public knowledge bases, and CTI reports~\cite{tamanna2026adversaries}. This observation motivates domain-specific CVE-based datasets for evaluating LLMs in structured threat intelligence generation.

\textbf{LLM-Based STIX and CTI Generation.}
Prior survey work shows that \gls{cti} extraction studies export structured threat knowledge in multiple formats, including knowledge graphs, \gls{stix}, Malware Information Sharing Platform, Open Indicators of Compromise, and domain-agnostic representations~\cite{rahman2023attackers}. Among these formats, \gls{stix} is widely used for representing \gls{cti} objects and relationships in a machine-readable form, and prior studies have explored how \gls{llms} and \gls{nlp} pipelines can generate or extract structured CTI from unstructured sources. For example, Time for aCTIon~\cite{siracusano2023time} evaluates GPT-3.5-turbo for converting real-world CTI reports into structured STIX information. Similarly, CTI-GEN~\cite{papoutsis2025cti} uses GPT-4o with multi-stage prompt engineering to generate STIX 2.1 objects and relationships from textual reports and manually constructed attack scenarios. In contrast, STIXnet~\cite{marchiori2023stixnet} focuses on \gls{sdo} through a hybrid pipeline rather than LLM evaluation. More recently, MAD-CTI~\cite{shah2025mad} evaluates GPT-4o mini, GPT-4o, and Claude 3.5 in a multi-agent framework for dark-web CTI analysis. Together, these studies provide evidence that LLMs and extraction can support structured CTI generation, but CAV-related CVE descriptions to STIX generation with open-weight LLMs evaluation remains underexplored.

\textbf{CWE and MITRE ATT\&CK Technique Analysis.}
Prior work also examines CWE classification and Mitre attack technique extraction from vulnerability or threat descriptions. For CWE-oriented analysis, CTIBench~\cite{alam2024ctibench} includes CVE-to-CWE prediction and CTI-based MITRE ATT\&CK technique extraction tasks. Similarly, Text2Weak~\cite{simonetto2024text2weak} focuses on mapping CVE descriptions to CWE categories. In the autonomous vehicle domain, Haque et al.~\cite{haque2025security} analyze autonomous vehicle software stacks and identify recurring CWE patterns and third-party library risks in platforms such as Autoware, Apollo, and openpilot. In parallel, several studies focus on MITRE ATT\&CK and TTP extraction from unstructured CTI. For example, TTPDrill~\cite{husari2017ttpdrill} extracts threat actions from unstructured CTI sources and maps them to tactics, techniques, and kill-chain phases. ThreatPilot~\cite{xu2024threatpilot} extracts Tactics, Techniques, and Procedures from CTI reports and supports Sigma rule generation. TIEF~\cite{joy2025threat} extracts TTPs from threat reports and represents attack behaviors in STIX format. Together, these studies support CWE prediction and ATT\&CK/TTP extraction, but they do not focus on CWE and MITRE ATT\&CK co-occurrence within CAV-focused ground-truth STIX data.

Prior work provides CTI extraction pipelines, LLM-based STIX generation methods, CVE understanding datasets, and ATT\&CK/TTP extraction techniques. Our work differs by focusing on CAV-related CVE descriptions, constructing a CAV-focused CVE-to-STIX dataset, evaluating open-weight LLMs for \gls{sdo} and \gls{sro} generation, and analyzing CWE and MITRE ATT\&CK mapping in ground-truth STIX data.

\section{Methodology}
\label{sec:methodology}
We discuss the methodology in this section.
\subsection{CAV-STIXGen Dataset Construction}
\label{subsec:cav_stixgen_construction}

We construct the \textit{CAV-STIXGen} dataset to address RQ1. 
\textbf{Step 1: Keyword construction.}
We built a domain-specific keyword to collect CAV-related vulnerabilities, given in Table~\ref{tab:unique_keywords} by asking \mycomment{}{ChatGPT, Copilot, Gemini, and Perplexity}{ChatGPT\footnote{\url{https://chatgpt.com/}}, Copilot\footnote{\url{https://copilot.microsoft.com/}}, Gemini\footnote{\url{https://gemini.google.com/app}}, and Perplexity\footnote{\url{https://www.perplexity.ai/}}}. The first and third authors then review, merge, and filter the generated keywords to remove duplicate and weakly related keywords.

\begin{table}[h!]
\scriptsize
\centering
\caption{Unique Keyword Categories}
\label{tab:unique_keywords}
\begin{tabular}{p{0.15\linewidth} p{0.75\linewidth}}
\hline
\textbf{Category} & \textbf{Unique Keywords} \\
\hline
Types 
& Connected vehicle, Automotive, \gls{ev} \\
\hline
Manufacturers 
& Audi, Tesla, BMW, Toyota, Mercedes-Benz, Ford, General Motors, Hyundai, Kia, Nissan, Volkswagen, BYD \\
\hline
Hardware and Systems 
& GPS in vehicle, Inertial Measurement Unit (IMU), Steering, Throttle, Controller Area Network (CAN) in vehicle \\
\hline
Connectivity 
& Vehicle-to-Vehicle (V2V), Vehicle-to-Cloud (V2C), Over-the-Air (OTA) updates \\
\hline
\end{tabular}
\end{table}

\textbf{Step 2: CVE Collection from NVD.}
We searched with the keywords the \gls{nvd} through REST API calls. The search returns 445 candidate CVE records. We then remove duplicates and retain 368 records.

\textbf{Step 3: Relevance filtering.} The first and second authors assess each CVE record using inclusion and exclusion criteria as follows. We include CVEs related to vehicle components, \gls{ev} charging, telematics, infotainment, Electronic Control Units (ECUs), CAN, Vehicle-to-Everything (V2X), sensors, and automotive infrastructure. We exclude CVEs with vulnerabilities affecting generic websites, sales platforms, administrative systems, or non-vehicle-specific software. The annotators achieve Cohen's $\kappa$ \cite{cohen1960coefficient} 94\% inter-rater agreement, and after resolving disagreements, we retain 183 unique records.

\textbf{Step 4: STIX Domain Objects and STIX Relationship Objects Annotation.}
We manually annotate \gls{sdo} and \gls{sro} for each retained CVE according to the STIX 2.1 specification~\cite{oasis_stix21}. We store the annotations in CSV files to enable review, correction, and validation. The annotations include the relevant STIX object types among the 19 \gls{sdo} types. We then annotate \gls{sro} by linking each source object to a target object using the relevant relationship types. We assign each object and relationship a unique identifier, denoted as \(E_i\) and \(R_i\), respectively, where \(i\) is a natural number.

\textbf{Step 5: CWE and MITRE ATT\&CK Mapping.}
We map each CVE with corresponding \gls{cwe} and Mitre attack technique mappings. We collect \gls{cwe} information from the \gls{nvd} when available. We then manually verify each NVD-provided CWE mapping against the corresponding CVE description to ensure that the assigned weakness type matches the vulnerability behavior. When the \gls{nvd} does not provide a CWE mapping, such as for CVE-2024-51074\footnote{\url{https://nvd.nist.gov/vuln/detail/CVE-2024-51074}}, we manually analyze the CVE description and assign the appropriate CWE based on the vulnerability behavior and root cause. We also map each extracted \texttt{attack-pattern} STIX object to the corresponding MITRE ATT\&CK techniques~\cite{mitre_attack}.

\textbf{Step 6: Validation and STIX JSON Conversion.}
After completing Step 5, we store all annotations in CSV files for inspection and correction. The first and second authors review the files, resolve disagreements, and revise ambiguous cases. After validation, we programmatically convert the CSV files into ground-truth STIX 2.1 JSON bundles.

\subsection{LLM Evaluation Using the CAV-STIXGen Dataset}
\label{subsec:llm_evaluation}

We use the \textit{CAV-STIXGen} dataset to evaluate LLM-based CVE-to-STIX generation. Each model receives a CAV-related CVE identifier and description as input and generates a STIX JSON output. We compare each generated output with the manually validated ground-truth STIX file for the same CVE. We evaluate each model under multiple prompting strategies and four temperature settings: \(0\), \(0.25\), \(0.75\), and \(1.0\), representing deterministic generation, low-variation generation, moderate-variation generation, and high-variation generation, respectively. The evaluation measures \gls{sdo} extraction, \gls{sro} construction, CWE mapping, and MITRE ATT\&CK technique mapping. We use micro-precision, micro-recall, and micro-F1 because CVE-to-STIX generation is an instance-level generation task, and micro-level aggregation captures overall correctness across all generated objects, relationships, CWE, and MITRE ATT\&CK mappings without over-weighting rare labels. For MITRE ATT\&CK mapping, we also report Match@1 and Match@All because a single CVE can be associated with multiple valid ATT\&CK techniques. Match@1 measures whether a model identifies at least one correct technique, while Match@All measures whether the model identifies all techniques from the ground-truth for that CVE.

We evaluate 11 open-weight LLMs across four model categories. \textbf{General-purpose models:} Gemma-3-4B, Gemma-4-31B, Microsoft Phi-4, and Qwen-3.5-9B. \textbf{Cybersecurity-specialized models:} Lily-CyberSecurity-7B and CyberSec-Qwen3-DeepSeekv1. \textbf{Code-oriented models:} Codestral-22B and Qwen3-Coder-30B. \textbf{Large-capacity open-weight models:} GPT-OSS-20B, GPT-OSS-120B, and LLaMA-3.3-70B. We evaluate a multi-agent configuration with Gemma-4-31B and Codestral to examine whether role-specific decomposition improves generation compared with single-model prompting. For space, tables use shortened model names where needed.\footnote{\textit{Note:} For space, Microsoft Phi-4 is shown as Phi-4, Qwen-3-Coder-30B as Qwen-Coder, Qwen-3.5-9B as Qwen-9B GPT-OSS-20B as GPT-20B, GPT-OSS-120B as GPT-120B, and LLaMA-3.3-70B as LLaMA-70B, Lily-CyberSecurity-7B as Lily-7B, CyberSec-Qwen3-DeepSeekv1 as CySec-Qwen and Codestral-22B as Codestral-22B.}

\begin{figure}[t]
    \centering
    \includegraphics[width=\columnwidth]{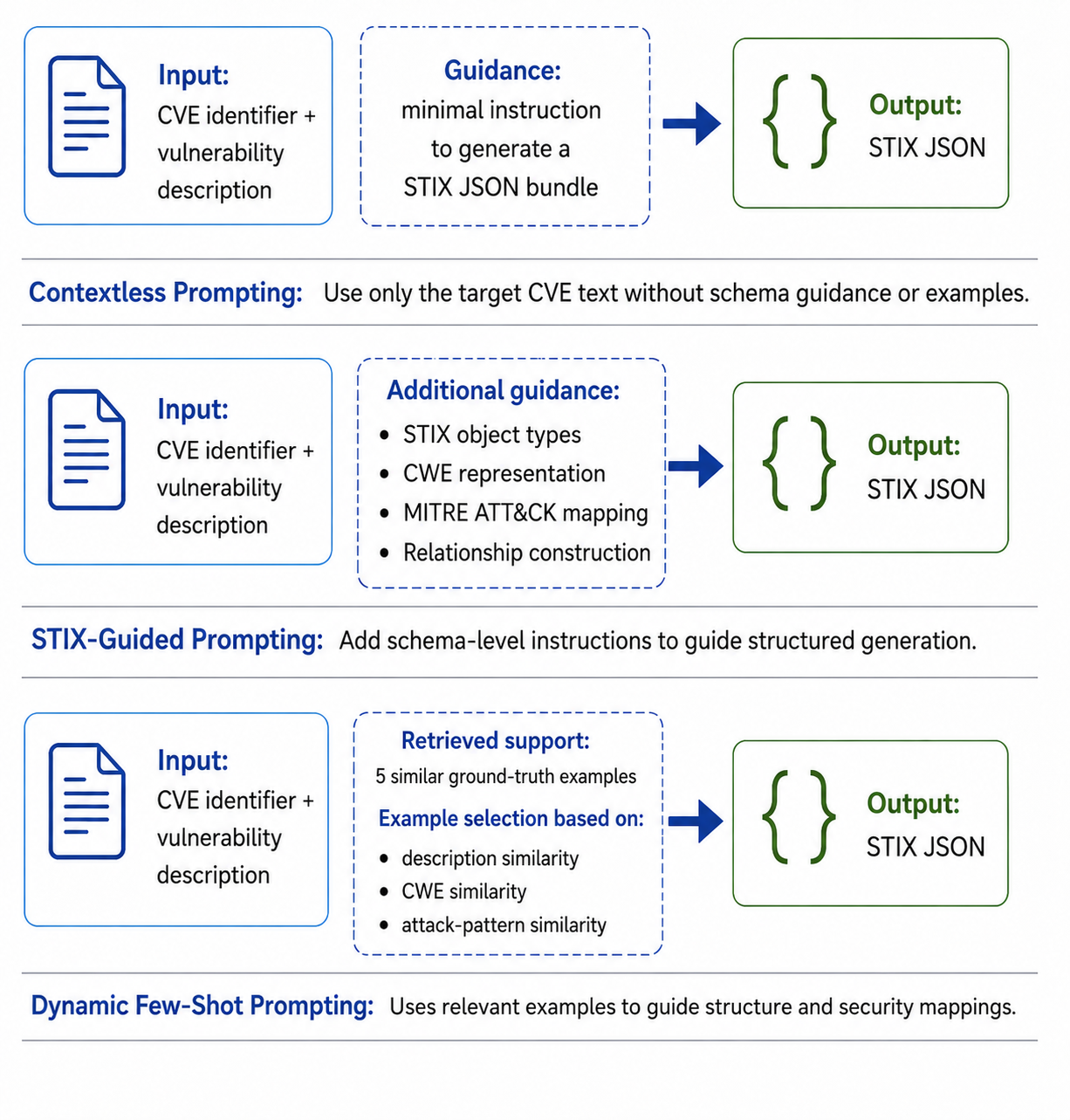}
    \caption{Prompting strategies for CVE-to-STIX generation.}
    \label{fig:prompting_strategies}
\end{figure}

Fig.~\ref{fig:prompting_strategies} summarizes the three prompting strategies used in our evaluation. \textbf{Contextless prompting} uses the CVE identifier and vulnerability description as inputs for generating a STIX JSON bundle. \textbf{STIX-guided prompting} uses the same input and adds task guidance for STIX object types, CWE representation, MITRE ATT\&CK mapping, and relationship construction. \textbf{Dynamic few-shot prompting} extends the guided setting with five similar ground-truth STIX examples using similarity score among description, CWE, and attack-pattern.

\subsection{CWE and MITRE ATT\&CK Technique Analysis}
\label{subsec:mitre_cooccurrence}
We perform the CWE and MITRE ATT\&CK Technique Analysis to address RQ3.

For CWE analysis, we examine the weakness mappings associated with each CAV-related CVE in the ground-truth data. We first count the frequency of each CWE category across the dataset to identify dominant weakness patterns in CAV-related vulnerabilities. We then compare the frequent CWE categories with the CWE Top 25 list~\cite{mitre_cwe_top25_2025} to contextualize whether the observed weaknesses align with widely recognized software security risks. This analysis helps characterize the weakness landscape of CAV vulnerabilities and explains which types of software flaws appear most often in the dataset.

Although MITRE ATT\&CK techniques are represented in STIX through \texttt{attack-pattern} SDOs, we analyze ATT\&CK mappings separately to measure whether these \texttt{attack-pattern} objects are correctly normalized to MITRE technique IDs. We also perform MITRE ATT\&CK technique co-occurrence analysis to identify recurring attack-behavior patterns in the ground-truth STIX data. Each CVE is treated as one transaction, and the MITRE ATT\&CK techniques extracted from the \texttt{attack-pattern} STIX objects are treated as transaction items. We extract MITRE technique IDs from the \texttt{external\_references} field of each \texttt{attack-pattern} object. We then apply frequent itemset mining using FP-Growth with a minimum support threshold of 0.02 to identify frequent individual techniques and technique combinations. We also generate association rules using a minimum confidence threshold of 0.30 to measure directional associations between techniques. Finally, we construct a pairwise co-occurrence network from the extracted techniques, where each node represents a MITRE technique, node size represents technique frequency, and edge weight represents the number of CVEs in which two techniques appear together.

\subsection{Experimental Setup}
\label{subsec:experimental_setup}

We conducted the CVE-to-STIX generation experiments using a local inference server and a development machine. The inference server ran \textbf{LM Studio version 0.4.8 (Build 1)} on Ubuntu and hosted the open-weight LLMs through the LM Studio OpenAI-compatible API. The server used four \textbf{NVIDIA RTX 6000 Ada Generation} GPUs with 48 GB of VRAM each, totaling 192 GB of GPU memory, with \textbf{NVIDIA Driver version 580.126.09} and \textbf{CUDA 13.0}. 

\section{\mycomment{Evaluation}{Results}}
We present the results for RQ1-RQ3 in this section.

\newcommand{\best}[1]{\cellcolor{green!10}{\textbf{#1}}}
\begin{table*}[!t]
\centering
\caption{CVE-to-STIX generation performance across prompting strategies}
\label{tab:all_task_compact}
\tiny
\setlength{\tabcolsep}{2.0pt}
\renewcommand{\arraystretch}{1.05}

\begin{tabular*}{\textwidth}{@{\extracolsep{\fill}}llccc|ccc|ccc|ccc|ccc@{}}
\hline
\multirow{2}{*}{\textbf{Model}} 
& \multirow{2}{*}{\textbf{T}} 
& \multicolumn{3}{c|}{\textbf{\gls{sdo}}} 
& \multicolumn{3}{c|}{\textbf{\gls{sro}}} 
& \multicolumn{3}{c|}{\textbf{CWE}} 
& \multicolumn{3}{c|}{\textbf{Attack Match@1}} 
& \multicolumn{3}{c}{\textbf{Attack Match@All}} \\
\cline{3-17}
& 
& \textbf{C} & \textbf{SG} & \textbf{DFS}
& \textbf{C} & \textbf{SG} & \textbf{DFS}
& \textbf{C} & \textbf{SG} & \textbf{DFS}
& \textbf{C} & \textbf{SG} & \textbf{DFS}
& \textbf{C} & \textbf{SG} & \textbf{DFS} \\
\hline

\multirow{4}{*}{\textbf{Gemma-3-4B}}
& \textbf{0.00} & 0.60 & 0.84 & 0.91 & 0.00 & 0.03 & 0.31 & 0.00 & 0.11 & 0.95 & 0.02 & 0.17 & 0.37 & 0.02 & 0.04 & 0.16 \\
& \textbf{0.25} & 0.59 & 0.85 & 0.91 & 0.00 & 0.03 & 0.31 & 0.00 & 0.11 & 0.95 & 0.01 & 0.16 & 0.39 & 0.01 & 0.04 & 0.16 \\
& \textbf{0.75} & 0.53 & 0.83 & 0.90 & 0.00 & 0.03 & 0.30 & 0.00 & 0.11 & 0.95 & 0.02 & 0.17 & 0.38 & 0.02 & 0.04 & 0.15 \\
& \textbf{1.00} & 0.50 & 0.77 & 0.89 & 0.00 & 0.01 & 0.28 & 0.00 & 0.10 & 0.93 & 0.01 & 0.14 & 0.37 & 0.01 & 0.03 & 0.14 \\
\hline

\multirow{4}{*}{\textbf{Gemma-4-31B}}
& \textbf{0.00} & 0.60 & 0.86 & 0.79 & 0.20 & 0.45 & 0.60 & 0.39 & 0.39 & 0.93 & 0.50 & 0.41 & 0.64 & \best{0.24} & 0.19 & 0.28 \\
& \textbf{0.25} & 0.60 & 0.86 & 0.80 & 0.20 & 0.44 & 0.61 & 0.38 & 0.37 & 0.95 & \best{0.52} & 0.41 & \best{0.68} & 0.23 & 0.18 & \best{0.33} \\
& \textbf{0.75} & 0.60 & 0.88 & 0.76 & 0.20 & 0.46 & 0.60 & 0.38 & 0.38 & 0.92 & 0.47 & 0.42 & 0.62 & \best{0.24} & \best{0.20} & 0.28 \\
& \textbf{1.00} & 0.60 & 0.88 & 0.78 & \best{0.21} & 0.47 & 0.61 & 0.37 & 0.37 & 0.94 & 0.43 & 0.42 & 0.66 & 0.22 & 0.18 & \best{0.33} \\
\hline

\multirow{4}{*}{\textbf{Phi-4}}
& \textbf{0.00} & 0.65 & 0.89 & \best{0.94} & 0.01 & 0.40 & 0.55 & 0.64 & \best{0.89} & 0.94 & 0.14 & 0.23 & 0.53 & 0.03 & 0.08 & 0.22 \\
& \textbf{0.25} & 0.65 & 0.89 & \best{0.94} & 0.02 & 0.38 & 0.55 & \best{0.65} & \best{0.89} & 0.94 & 0.12 & 0.22 & 0.51 & 0.03 & 0.09 & 0.22 \\
& \textbf{0.75} & 0.60 & 0.87 & \best{0.94} & 0.03 & 0.38 & 0.53 & 0.58 & 0.85 & 0.93 & 0.09 & 0.15 & 0.52 & 0.01 & 0.07 & 0.20 \\
& \textbf{1.00} & 0.59 & 0.86 & \best{0.94} & 0.03 & 0.35 & 0.53 & 0.56 & 0.84 & 0.94 & 0.06 & 0.18 & 0.49 & 0.01 & 0.04 & 0.19 \\
\hline

\multirow{4}{*}{\textbf{Qwen-9B}}
& \textbf{0.00} & 0.00 & 0.14 & 0.80 & 0.00 & 0.07 & 0.50 & 0.00 & 0.08 & 0.81 & 0.00 & 0.04 & 0.52 & 0.00 & 0.01 & 0.23 \\
& \textbf{0.25} & 0.00 & 0.17 & 0.75 & 0.00 & 0.08 & 0.47 & 0.00 & 0.09 & 0.76 & 0.00 & 0.04 & 0.45 & 0.00 & 0.02 & 0.22 \\
& \textbf{0.75} & 0.00 & 0.15 & 0.75 & 0.00 & 0.05 & 0.44 & 0.00 & 0.07 & 0.77 & 0.00 & 0.05 & 0.47 & 0.00 & 0.02 & 0.23 \\
& \textbf{1.00} & 0.00 & 0.11 & 0.78 & 0.00 & 0.04 & 0.46 & 0.00 & 0.09 & 0.80 & 0.00 & 0.02 & 0.49 & 0.00 & 0.01 & 0.22 \\
\hline

\multirow{4}{*}{\textbf{Lily-7B}}
& \textbf{0.00} & 0.00 & 0.83 & 0.87 & 0.00 & 0.00 & 0.37 & 0.00 & 0.11 & 0.96 & 0.00 & 0.07 & 0.28 & 0.00 & 0.04 & 0.12 \\
& \textbf{0.25} & 0.00 & 0.78 & 0.87 & 0.00 & 0.00 & 0.41 & 0.00 & 0.10 & 0.95 & 0.00 & 0.07 & 0.29 & 0.00 & 0.03 & 0.11 \\
& \textbf{0.75} & 0.00 & 0.69 & 0.83 & 0.00 & 0.00 & 0.31 & 0.00 & 0.10 & 0.90 & 0.00 & 0.02 & 0.20 & 0.00 & 0.00 & 0.08 \\
& \textbf{1.00} & 0.00 & 0.62 & 0.78 & 0.00 & 0.00 & 0.25 & 0.00 & 0.11 & 0.82 & 0.00 & 0.05 & 0.13 & 0.00 & 0.02 & 0.05 \\
\hline

\multirow{4}{*}{\textbf{CySec-Qwen}}
& \textbf{0.00} & 0.52 & 0.28 & 0.86 & 0.01 & 0.15 & 0.39 & 0.00 & 0.05 & 0.90 & 0.02 & 0.09 & 0.42 & 0.01 & 0.02 & 0.19 \\
& \textbf{0.25} & 0.52 & 0.20 & 0.84 & 0.00 & 0.09 & 0.38 & 0.02 & 0.02 & 0.89 & 0.02 & 0.07 & 0.39 & 0.01 & 0.02 & 0.19 \\
& \textbf{0.75} & 0.41 & 0.08 & 0.71 & 0.00 & 0.03 & 0.28 & 0.00 & 0.01 & 0.73 & 0.02 & 0.01 & 0.24 & 0.01 & 0.00 & 0.10 \\
& \textbf{1.00} & 0.32 & 0.06 & 0.61 & 0.00 & 0.02 & 0.20 & 0.00 & 0.02 & 0.61 & 0.01 & 0.01 & 0.14 & 0.00 & 0.01 & 0.09 \\
\hline

\multirow{4}{*}{\textbf{Codestral-22B}}
& \textbf{0.00} & 0.57 & 0.89 & \best{0.94} & 0.15 & 0.22 & 0.55 & 0.11 & 0.16 & \best{0.99} & 0.15 & 0.29 & 0.46 & 0.07 & 0.14 & 0.19 \\
& \textbf{0.25} & 0.56 & 0.88 & 0.93 & 0.15 & 0.20 & 0.54 & 0.10 & 0.15 & 0.98 & 0.11 & 0.26 & 0.45 & 0.04 & 0.10 & 0.19 \\
& \textbf{0.75} & 0.49 & 0.88 & 0.92 & 0.13 & 0.17 & 0.53 & 0.12 & 0.18 & 0.98 & 0.07 & 0.22 & 0.46 & 0.03 & 0.09 & 0.19 \\
& \textbf{1.00} & 0.48 & 0.86 & 0.88 & 0.13 & 0.17 & 0.50 & 0.07 & 0.19 & 0.96 & 0.08 & 0.21 & 0.42 & 0.02 & 0.09 & 0.18 \\
\hline

\multirow{4}{*}{\textbf{Qwen-Coder}}
& \textbf{0.00} & 0.54 & 0.84 & \best{0.94} & 0.03 & 0.38 & 0.62 & 0.01 & 0.03 & \best{0.99} & 0.23 & 0.27 & 0.67 & 0.11 & 0.12 & 0.26 \\
& \textbf{0.25} & 0.55 & 0.85 & \best{0.94} & 0.04 & 0.36 & 0.62 & 0.02 & 0.01 & \best{0.99} & 0.20 & 0.28 & 0.64 & 0.10 & 0.13 & 0.27 \\
& \textbf{0.75} & 0.54 & 0.84 & 0.93 & 0.04 & 0.37 & 0.62 & 0.01 & 0.02 & 0.98 & 0.19 & 0.23 & 0.64 & 0.09 & 0.08 & 0.27 \\
& \textbf{1.00} & 0.55 & 0.84 & \best{0.94} & 0.04 & 0.37 & \best{0.63} & 0.03 & 0.01 & \best{0.99} & 0.17 & 0.25 & 0.67 & 0.09 & 0.13 & 0.28 \\
\hline

\multirow{4}{*}{\textbf{GPT-20B}}
& \textbf{0.00} & 0.60 & 0.85 & 0.84 & 0.10 & 0.41 & 0.51 & 0.12 & 0.27 & 0.95 & 0.17 & 0.28 & 0.56 & 0.07 & 0.13 & 0.29 \\
& \textbf{0.25} & 0.61 & 0.85 & 0.81 & 0.12 & 0.40 & 0.48 & 0.12 & 0.25 & 0.93 & 0.20 & 0.26 & 0.54 & 0.09 & 0.11 & 0.25 \\
& \textbf{0.75} & 0.60 & 0.87 & 0.77 & 0.11 & 0.39 & 0.46 & 0.11 & 0.23 & 0.88 & 0.21 & 0.23 & 0.46 & 0.07 & 0.10 & 0.24 \\
& \textbf{1.00} & 0.57 & 0.84 & 0.75 & 0.09 & 0.37 & 0.45 & 0.12 & 0.17 & 0.85 & 0.18 & 0.20 & 0.49 & 0.08 & 0.07 & 0.24 \\
\hline

\multirow{4}{*}{\textbf{LLaMA-70B}}
& \textbf{0.00} & 0.68 & 0.88 & \best{0.94} & 0.13 & 0.38 & 0.51 & 0.09 & 0.35 & \best{0.99} & 0.17 & 0.12 & 0.61 & 0.08 & 0.03 & 0.27 \\
& \textbf{0.25} & \best{0.69} & 0.88 & \best{0.94} & 0.12 & 0.38 & 0.51 & 0.10 & 0.35 & \best{0.99} & 0.24 & 0.14 & 0.61 & 0.12 & 0.06 & 0.26 \\
& \textbf{0.75} & 0.66 & 0.88 & \best{0.94} & 0.13 & 0.39 & 0.51 & 0.13 & 0.31 & 0.98 & 0.17 & 0.13 & 0.59 & 0.09 & 0.06 & 0.27 \\
& \textbf{1.00} & 0.67 & 0.88 & \best{0.94} & 0.12 & 0.38 & 0.50 & 0.13 & 0.31 & \best{0.99} & 0.15 & 0.14 & 0.57 & 0.08 & 0.06 & 0.26 \\
\hline

\multirow{4}{*}{\textbf{GPT-120B}}
& \textbf{0.00} & 0.59 & \best{0.90} & 0.83 & 0.20 & \best{0.49} & 0.52 & 0.31 & 0.35 & 0.98 & 0.34 & \best{0.44} & 0.57 & 0.15 & \best{0.20} & 0.27 \\
& \textbf{0.25} & 0.60 & \best{0.90} & 0.83 & 0.18 & 0.48 & 0.53 & 0.28 & 0.33 & \best{0.99} & 0.39 & 0.37 & 0.58 & 0.17 & 0.18 & 0.29 \\
& \textbf{0.75} & 0.60 & \best{0.90} & 0.82 & 0.17 & 0.48 & 0.52 & 0.30 & 0.29 & \best{0.99} & 0.39 & 0.39 & 0.61 & 0.16 & 0.17 & 0.30 \\
& \textbf{1.00} & 0.60 & \best{0.90} & 0.83 & 0.18 & 0.47 & 0.51 & 0.27 & 0.30 & 0.98 & 0.36 & 0.41 & 0.57 & 0.14 & 0.18 & 0.27 \\
\hline

\hline
\end{tabular*}

\vspace{1mm}
\footnotesize{
\textit{Note:} C = Contextless, SG = STIX-Guided, DFS = Dynamic Few-Shot. 
Objects, Relationship, and CWE are F1 scores. 
Match@1 and Match@All are MITRE ATT\&CK mapping scores. Green box represents the largest value in a column.
}
\end{table*}

\subsection{RQ1: Construction \mycomment{and Characterization}{}of CAV-STIXGen Dataset}

\begin{table}[t]
\centering
\caption{Multi-Agent \gls{sdo} and \gls{sro} Extraction}
\label{tab:multiagent_entity_relationship}
\scriptsize
\setlength{\tabcolsep}{5pt}
\begin{tabular}{llccc|ccc}
\hline
\textbf{Model} & \textbf{Temp.} 
& \multicolumn{3}{c|}{\textbf{Objects}} 
& \multicolumn{3}{c}{\textbf{Relationship}} \\
\cline{3-8}
& & \textbf{P} & \textbf{R} & \textbf{F1}
& \textbf{P} & \textbf{R} & \textbf{F1} \\
\hline
MultiAgent (Gemma-4-31B) 
& 0.75 & 0.94 & 0.88 & 0.91 & 0.38 & 0.40 & 0.39 \\
MultiAgent (Codestral-22B) 
& 0    & 0.90 & 0.83 & 0.86 & 0.50 & 0.38 & 0.43 \\
\hline
\end{tabular}
\end{table}

\textbf{Ground-Truth STIX Bundle Characteristics.}
Table~\ref{tab:dataset_composition} summarizes the composition of \textit{CAV-STIXGen}. The dataset starts from 445 candidate CVE records collected from NVD and retains 183 CAV-related CVEs after relevance filtering. The final ground-truth dataset contains \textbf{1,383 \gls{sdo} instances}, \textbf{1,395 \gls{sro} instances}, \textbf{211 CWE mappings}, and \textbf{294 MITRE ATT\&CK mappings}. The dataset covers \textbf{10 out of 19 \gls{sdo} types}: \texttt{infrastructure}, \texttt{attack-pattern}, \texttt{vulnerability}, \texttt{identity}, \texttt{threat-actor}, \texttt{note}, \texttt{tool}, \texttt{indicator}, \texttt{observed-data}, and \texttt{malware}. The most frequent \gls{sdo} types are \texttt{infrastructure}, \texttt{attack-pattern}, \texttt{vulnerability}, \texttt{identity}, and \texttt{threat-actor}, indicating that the dataset captures affected assets, attack behaviors, weaknesses, vendors, and attacker roles. The most frequent \gls{sro} types are \texttt{exploits}, \texttt{targets}, \texttt{uses}, \texttt{affects}, \texttt{part-of}, \texttt{produces}, \texttt{enables}, \texttt{has-version}, \texttt{deployed-in}, and \texttt{used-by}. These relationship types show how attack behaviors exploit vulnerabilities, target systems, use resources, affect infrastructures, and connect software components to CAV-related assets. On average, each CVE contains \textbf{7.56 \gls{sdo} instances}, \textbf{7.62 \gls{sro} instances}, \textbf{1.15 CWE mappings}, and \textbf{1.61 MITRE ATT\&CK mappings}.

\begin{table}[t]
\centering
\caption{Statistics of CAV-STIXGen Dataset Composition}
\label{tab:dataset_composition}
\scriptsize
\begin{tabular}{lc}
\hline
\textbf{Dataset Characteristic} & \textbf{Value} \\
\hline
Candidate CVE records collected from NVD & 445 \\
Retained CAV-related CVEs & 183 \\
Data collection period & 2012--2025 \\
Most frequent CVE year & 2023, 61 CVEs \\
Total STIX objects & 1,383 \\
Total STIX relationships & 1,395 \\
Average STIX objects per CVE & 7.56 \\
Average STIX relationships per CVE & 7.62 \\
STIX object types covered & 10 of 19 \\
Most frequent STIX object type & Infrastructure, 538 \\
Total CWE mappings & 211 \\
Average CWE mappings per CVE & 1.15 \\
Most frequent CWE & CWE-787, CWE-20 \\
Top-10 CWEs appearing in CWE Top 25 & 5 of 10 \\
Total MITRE ATT\&CK mappings & 294 \\
Average MITRE ATT\&CK mappings per CVE & 1.61 \\
Most frequent MITRE ATT\&CK technique & T1499, 48 \\
Most frequent ATT\&CK co-occurring pair & T1203--T1499, 10 CVEs \\
Annotation format & CSV \& JSON \\
\hline
\end{tabular}
\end{table}

\subsection{RQ2: Open-Weight LLM Evaluation}
\label{subsec:rq2_llm_evaluation}

RQ2 evaluates how effectively open-weight LLMs generate STIX representations from CAV-related CVE descriptions under different prompting strategies, shown in Table~\ref{tab:all_task_compact}. We compare contextless prompting, STIX-guided prompting, and dynamic few-shot prompting across five evaluation dimensions: STIX object extraction, STIX relationship extraction, CWE mapping, MITRE ATT\&CK Match@1, and Match@All. \mycomment{Table~\ref{tab:all_task_compact} summarizes the results across models, temperatures, prompting strategies, and tasks.}{}

\subsubsection{Overall Prompting Strategy Comparison}

The results show a performance difference across prompting strategies. Contextless prompting provides weaker performance because models receive only minimal task instructions and \mycomment{must}{}infer STIX without examples. STIX-guided prompting improves STIX generation by providing explicit instructions for \gls{sdo} types, \gls{sro}, CWE references, and MITRE ATT\&CK mappings. Dynamic few-shot prompting gives the strongest overall results because target-relevant examples help models produce more complete STIX bundles. For example, Phi-4 achieves an \gls{sdo} extraction F1 score of 0.94 under dynamic few-shot, while Qwen-Coder achieves a relationship extraction F1 score of 0.63. The results indicate that \gls{sdo} examples improves \gls{llms} performance on STIX generation\mycomment{structure}{}, while dynamically selected examples improve both structure and semantic mapping.

\subsubsection{Performance Across CVE-to-STIX Tasks}

In this task, STIX object extraction obtains the highest scores across most models. Under dynamic few-shot prompting, Phi-4, LLaMA-70B, Codestral-22B, and Qwen-Coder reach object extraction F1 scores around 0.94. Relationship extraction obtains lower scores because the task requires models to identify the correct source object, target object, and relationship type. Qwen-Coder achieves the best relationship extraction result, with an F1 score of 0.63 under dynamic few-shot prompting.

CWE mapping shows improvement when similar examples are provided. Under dynamic few-shot prompting, Codestral-22B, Qwen-Coder, LLaMA-70B, and GPT-120B each achieve CWE F1 scores close to 0.99. This result indicates that example-guided prompting helps models connect vulnerability descriptions with weakness categories and normalize CWE identifiers more accurately than contextless prompting.

MITRE ATT\&CK mapping requires richer attack-behavior reasoning than CWE mapping. Match@1 improves under dynamic few-shot prompting, which indicates that models can often identify at least one correct ATT\&CK technique. Match@All remains lower because a single CVE can involve multiple valid ATT\&CK techniques. 

\subsubsection{Model-Level Findings}

Model performance varies across subtasks. LLaMA-70B, Phi-4, Codestral-22B, Qwen-Coder, and GPT-120B perform strongly for STIX object extraction under guided or few-shot prompting. Qwen-Coder achieves the strongest relationship extraction score, followed by Gemma-4-31B, Codestral-22B, Phi-4, and GPT-120B under dynamic few-shot prompting. For CWE mapping, dynamic few-shot prompting produces the most consistent gains across larger and code-oriented models. For MITRE ATT\&CK mapping, Gemma-4-31B achieves the highest Match@1 score of 0.68, followed by Qwen-Coder with 0.67. These results suggest that no single model is best for every subtask; object extraction, relationship construction, CWE mapping, and ATT\&CK mapping require different reasoning capabilities.

\subsubsection{\gls{sdo} Type Analysis}

Table~\ref{tab:entity_f1_selected_objects_one_column} shows that performance differs across STIX object types. Vulnerability extraction is consistently strong because CVE descriptions explicitly describe the vulnerability. For example, Phi-4, Codestral-22B, Qwen-Coder, and LLaMA-70B reach vulnerability F1 scores close to 0.99 in at least one prompting setting. Other object types require more contextual inference. Threat-actor, infrastructure, and identity\mycomment{, and note}{} objects are missed under contextless prompting, but their extraction improves with STIX-guided or dynamic few-shot prompting. Dynamic few-shot prompting is especially effective for implicit object types because similar examples show how contextual information should be represented as STIX objects.

\begin{table}[t]
\centering
\caption{STIX Domain Object Type F1 Scores}
\label{tab:entity_f1_selected_objects_one_column}
\scriptsize
\setlength{\tabcolsep}{2.5pt}
\resizebox{\columnwidth}{!}{
\begin{tabular}{llccccc}
\hline
\textbf{Model} & \textbf{Prompt} 
& \textbf{AP} & \textbf{Vul.} & \textbf{TA} & \textbf{ID} & \textbf{Infra.} \\
\hline

\multirow{3}{*}{Gemma-3-4B}
& C (0)      & 0.65 & 0.88 & 0.53 & 0.34 & 0.05 \\
& SG (0.25)  & 0.78 & 0.98 & 0.82 & 0.42 & 0.59 \\
& DFS (0)    & 0.75 & 0.98 & 0.82 & 0.84 & 0.68 \\
\hline

\multirow{3}{*}{Gemma-4-31B}
& C (0)      & 0.75 & 0.99 & 0.00 & 0.14 & 0.03 \\
& SG (1.0)   & 0.80 & 0.96 & 0.82 & 0.77 & 0.62 \\
& DFS (0.25) & 0.83 & 0.95 & 0.89 & 0.80 & \best{0.83} \\
\hline

\multirow{3}{*}{Phi-4}
& C (0)      & 0.72 & 0.98 & 0.04 & 0.61 & 0.00 \\
& SG (0.25)  & 0.76 & \best{0.99} & 0.62 & 0.82 & 0.49 \\
& DFS (0.25) & 0.78 & \best{0.99} & 0.91 & 0.82 & 0.73 \\
\hline

\multirow{2}{*}{Qwen-9B}
& SG (0.25)  & 0.20 & 0.30 & 0.08 & 0.26 & 0.11 \\
& DFS (0)    & 0.67 & 0.83 & 0.80 & 0.69 & 0.71 \\
\hline

\multirow{2}{*}{Lily-7B}
& SG (0)     & 0.74 & 0.99 & 0.59 & 0.62 & 0.34 \\
& DFS (0.25) & 0.55 & 0.98 & 0.84 & 0.81 & 0.73 \\
\hline

\multirow{3}{*}{CySec-Qwen}
& C (0.25)   & 0.67 & 0.94 & 0.00 & 0.01 & 0.00 \\
& SG (0)     & 0.35 & 0.50 & 0.04 & 0.34 & 0.21 \\
& DFS (0)    & 0.73 & 0.94 & 0.67 & 0.78 & 0.57 \\
\hline

\multirow{3}{*}{Codestral-22B}
& C (0)      & 0.70 & 0.99 & 0.00 & 0.00 & 0.00 \\
& SG (0)     & 0.80 & \best{0.99} & 0.75 & 0.80 & 0.46 \\
& DFS (0)    & 0.81 & \best{0.99} & \best{0.92} & \best{0.84} & 0.75 \\
\hline

\multirow{3}{*}{Qwen-Coder}
& C (1.0)    & 0.74 & \best{0.99} & 0.00 & 0.21 & 0.00 \\
& SG (0.25)  & 0.77 & 0.99 & 0.21 & 0.79 & 0.53 \\
& DFS (0.25) & \best{0.87} & \best{0.99} & 0.90 & 0.84 & 0.81 \\
\hline

\multirow{3}{*}{GPT-20B}
& C (0.25)   & 0.65 & 0.99 & 0.00 & 0.56 & 0.00 \\
& SG (0.75)  & 0.77 & 0.98 & 0.64 & 0.81 & 0.56 \\
& DFS (0)    & 0.77 & 0.95 & 0.85 & 0.79 & 0.78 \\
\hline

\multirow{3}{*}{LLaMA-70B}
& C (0.25)   & 0.74 & \best{0.99} & 0.00 & 0.74 & 0.00 \\
& SG (0.25)  & 0.78 & \best{0.99} & 0.58 & 0.79 & 0.51 \\
& DFS (1.0)  & 0.79 & \best{0.99} & 0.89 & 0.82 & 0.76 \\
\hline

\multirow{3}{*}{GPT-120B}
& C (0.25)   & 0.73 & 0.99 & 0.00 & 0.22 & 0.00 \\
& SG (0)     & 0.81 & 0.99 & 0.84 & 0.79 & 0.54 \\
& DFS (1.0)  & 0.79 & 0.99 & 0.88 & 0.82 & 0.82 \\
\hline

\end{tabular}
}
\vspace{1mm}

\parbox{\columnwidth}{
\footnotesize{\textit{Note:} AP = Attack Pattern, Vul. = Vulnerability, TA = Threat Actor, ID = Identity, Infra. = Infrastructure, C = Contextless, SG = STIX-Guided, DFS = Dynamic Few-Shot. Temperature values are shown in parentheses.}
}
\end{table}

\subsubsection{STIX Relationship Object Type Analysis}

Table~\ref{tab:relationship_f1_selected_types} reports F1 scores for the top 10 \gls{sro} types across all evaluated \gls{llms}. The analysis focuses on these frequent relationship types because they represent the dominant edge patterns in the ground-truth STIX bundles. Relationship extraction is more complex than object extraction because models must identify the relevant STIX Domain Objects, determine the correct source and target objects, and assign the appropriate relationship type. Under contextless prompting, model outputs mainly capture explicit relationships such as \texttt{uses}, \texttt{targets}, and \texttt{exploits}. In contrast, semantically richer relationships such as \texttt{affects}, \texttt{part-of}, \texttt{produces}, \texttt{has-version}, \texttt{deployed-in}, and \texttt{used-by} often receive low F1 scores. STIX-guided prompting improves relationship extraction by providing examples on source-target linking and relationship semantics. Dynamic few-shot prompting provides the broadest relationship-type coverage because similar ground-truth examples demonstrate how vulnerabilities, attack patterns, infrastructures, and vendors should be connected in STIX. Overall, these results indicate that relationship extraction requires graph-level reasoning beyond object identification.

\subsubsection{Multi-Agent STIX Generation}
\label{subsubsec:multiagent_stix_generation}
The multi-agent configuration separates CVE-to-STIX generation into \gls{sdo} extraction, CWE mapping, MITRE ATT\&CK mapping, relationship construction, STIX JSON generation, and validation. Table~\ref{tab:multiagent_entity_relationship} summarizes the \gls{sdo} and \gls{sro} extraction performance of the Gemma-4-31B and Codestral-22B multi-agent configuration. We select Gemma-4-31B and Codestral-22B for the multi-agent setting because their single-prompt results show strong performance across \gls{sdo} extraction, relationship construction, and STIX generation. The results indicate that role-based decomposition can improve \gls{sdo} extraction for some models, but the improvement does not consistently transfer to \gls{sro} extraction. Relationship construction remains more complex because the configuration must connect the correct source and target objects through valid STIX relationship types.

\begin{table}[t]
\centering
\caption{STIX Relationship Object Type F1 Scores}
\label{tab:relationship_f1_selected_types}
\scriptsize
\setlength{\tabcolsep}{0.5pt}
\begin{tabular}{llcccccccccc}
\hline
\textbf{Model} & \textbf{Prompt} 
& \textbf{Exp.} & \textbf{Tar.} & \textbf{Use} & \textbf{Aff.} 
& \textbf{Part} & \textbf{Prod.} & \textbf{Enab.} 
& \textbf{Ver.} & \textbf{Dep.} & \textbf{U-by} \\
\hline

\multirow{2}{*}{Gemma-3-4B}
& SG (0)  & 0.08 & 0.02 & 0.06 & 0.02 & 0.00 & 0.00 & 0.00 & 0.00 & 0.00 & 0.00 \\
& DFS (0) & 0.22 & 0.01 & 0.09 & 0.90 & 0.50 & 0.08 & 0.00 & 0.03 & 0.18 & \best{0.36} \\
\hline

\multirow{3}{*}{Gemma-4-31B}
& C (1.0)   & 0.56 & 0.29 & 0.05 & 0.07 & 0.00 & 0.00 & 0.00 & 0.00 & 0.00 & 0.00 \\
& SG (1.0)  & \best{0.70} & 0.13 & 0.63 & 0.86 & 0.26 & 0.19 & \best{0.39} & 0.00 & 0.24 & 0.00 \\
& DFS (1.0) & 0.65 & 0.38 & 0.65 & 0.84 & \best{0.71} & \best{0.72} & 0.00 & \best{0.65} & 0.42 & 0.29 \\
\hline

\multirow{3}{*}{Phi-4}
& C (0.75) & 0.00 & 0.07 & 0.15 & 0.00 & 0.00 & 0.00 & 0.00 & 0.00 & 0.00 & 0.00 \\
& SG (0)   & 0.28 & \best{0.61} & 0.32 & 0.89 & 0.03 & 0.53 & 0.00 & 0.00 & 0.16 & 0.00 \\
& DFS (0)  & 0.56 & 0.44 & 0.45 & 0.91 & 0.58 & 0.68 & 0.00 & 0.43 & 0.44 & 0.10 \\
\hline

\multirow{2}{*}{Qwen-9B}
& SG (0.25) & 0.08 & 0.09 & 0.05 & 0.28 & 0.00 & 0.02 & 0.06 & 0.00 & 0.00 & 0.00 \\
& DFS (0)   & 0.56 & 0.31 & 0.56 & 0.75 & 0.53 & 0.56 & 0.00 & 0.61 & 0.21 & 0.00 \\
\hline

\multirow{1}{*}{Lily-7B}
& DFS (0.25) & 0.34 & 0.29 & 0.16 & 0.82 & 0.62 & 0.24 & 0.00 & 0.40 & 0.28 & 0.33 \\
\hline

\multirow{3}{*}{CySec-Qwen}
& C (0)   & 0.00 & 0.00 & 0.05 & 0.00 & 0.00 & 0.00 & 0.00 & 0.00 & 0.00 & 0.00 \\
& SG (0)  & 0.30 & 0.38 & 0.09 & 0.02 & 0.01 & 0.00 & 0.00 & 0.00 & 0.00 & 0.00 \\
& DFS (0) & 0.42 & 0.18 & 0.18 & 0.85 & 0.50 & 0.32 & 0.00 & 0.15 & 0.18 & 0.00 \\
\hline

\multirow{3}{*}{Codestral-22B}
& C (0)   & 0.01 & 0.15 & 0.63 & 0.00 & 0.00 & 0.00 & 0.00 & 0.00 & 0.00 & 0.00 \\
& SG (0)  & 0.62 & 0.42 & 0.52 & 0.77 & 0.01 & 0.33 & 0.02 & 0.00 & 0.00 & 0.00 \\
& DFS (0) & 0.58 & 0.39 & 0.65 & \best{0.92} & 0.53 & 0.58 & 0.00 & 0.34 & 0.35 & 0.12 \\
\hline

\multirow{3}{*}{Qwen-Coder}
& C (1.0)   & 0.14 & 0.07 & 0.02 & 0.00 & 0.00 & 0.00 & 0.00 & 0.00 & 0.00 & 0.00 \\
& SG (0)    & 0.67 & 0.25 & 0.07 & 0.74 & 0.07 & 0.33 & 0.07 & 0.00 & 0.00 & 0.00 \\
& DFS (1.0) & 0.68 & 0.60 & 0.60 & 0.91 & 0.68 & \best{0.72} & 0.02 & 0.54 & \best{0.49} & 0.32 \\
\hline

\multirow{3}{*}{GPT-20B}
& C (0.25) & 0.01 & 0.34 & 0.34 & 0.04 & 0.00 & 0.00 & 0.00 & 0.00 & 0.00 & 0.00 \\
& SG (0)   & 0.59 & 0.43 & 0.56 & 0.83 & 0.09 & 0.08 & 0.00 & 0.03 & 0.11 & 0.00 \\
& DFS (0)  & 0.52 & 0.42 & 0.65 & 0.83 & 0.61 & 0.28 & 0.00 & 0.43 & 0.38 & \best{0.36} \\
\hline

\multirow{3}{*}{LLaMA-70B}
& C (0.75)   & 0.13 & 0.50 & 0.09 & 0.01 & 0.00 & 0.00 & 0.00 & 0.00 & 0.00 & 0.00 \\
& SG (0.75)  & 0.66 & 0.14 & 0.37 & 0.91 & 0.01 & 0.12 & 0.00 & 0.00 & 0.00 & 0.00 \\
& DFS (0.75) & 0.55 & 0.28 & 0.40 & 0.91 & 0.66 & 0.45 & 0.00 & 0.25 & 0.27 & 0.33 \\
\hline

\multirow{3}{*}{GPT-120B}
& C (0)      & 0.27 & 0.27 & 0.53 & 0.01 & 0.00 & 0.00 & 0.00 & 0.00 & 0.00 & 0.00 \\
& SG (0)     & 0.67 & 0.30 & 0.62 & 0.90 & 0.15 & 0.63 & 0.07 & 0.00 & 0.00 & 0.00 \\
& DFS (0.25) & 0.53 & 0.48 & \best{0.67} & 0.89 & 0.66 & 0.54 & 0.00 & 0.53 & 0.41 & 0.25 \\
\hline

\end{tabular}
\vspace{1mm}

\parbox{\columnwidth}{
\footnotesize{\textit{Note:} Exp. = \texttt{exploits}, Tar. = \texttt{targets}, Use = \texttt{uses}, Aff. = \texttt{affects}, Part = \texttt{part-of}, Prod. = \texttt{produces}, Enab. = \texttt{enables}, Ver. = \texttt{has-version}, Dep. = \texttt{deployed-in}, U-by = \texttt{used-by}. }
}
\end{table}

\subsection{RQ3: CWE and MITRE ATT\&CK Technique Analysis}
\label{subsec:rq3_mitre_results}

Table~\ref{tab:top10_cwe_ground_truth} shows that \texttt{CWE-787}, \texttt{CWE-284}, and \texttt{CWE-20} are the most frequent weaknesses in our dataset. Five of the top 10 CWEs also appear in the 2025 CWE Top 25~\cite{mitre_cwe_top25_2025} list: \texttt{CWE-787}, \texttt{CWE-284}, \texttt{CWE-20}, \texttt{CWE-476}, and \texttt{CWE-863}. These weaknesses cover 57 of the 94 top-10 CWE occurrences, or 60.64\%. The overlap indicates that CAV-related vulnerabilities often involve widely recognized high-risk weakness categories, especially memory-safety, access-control, input-validation, and authorization weaknesses.

\begin{table}[t]
\centering
\caption{Top 10 CWE Categories in Dataset}
\label{tab:top10_cwe_ground_truth}
\scriptsize
\resizebox{\columnwidth}{!}{
\begin{tabular}{llcc}
\hline
\textbf{CWE} & \textbf{Name} & \textbf{Count} & \textbf{CWE Top 25} \\
\hline
CWE-787 & Out-of-bounds Write & 20 & Yes \\
CWE-284 & Improper Access Control & 14 & Yes \\
CWE-20  & Improper Input Validation & 12 & Yes \\
CWE-294 & Authentication Bypass by Capture-replay & 10 & No \\
CWE-287 & Improper Authentication & 9 & No \\
CWE-693 & Protection Mechanism Failure & 8 & No \\
CWE-190 & Integer Overflow or Wraparound & 6 & No \\
CWE-476 & NULL Pointer Dereference & 6 & Yes \\
CWE-863 & Incorrect Authorization & 5 & Yes \\
CWE-310 & Cryptographic Issues & 4 & No \\
\hline
\multicolumn{2}{l}{\textbf{Top-10 CWEs appearing in CWE Top 25}} & \multicolumn{2}{c}{\textbf{5/10}} \\
\multicolumn{2}{l}{\textbf{Occurrences covered by CWE Top 25}} & \multicolumn{2}{c}{\textbf{57/94 = 60.64\%}} \\
\hline
\end{tabular}
}
\end{table}

Table~\ref{tab:top10_mitre_attack_patterns} shows that the top 10 MITRE ATT\&CK techniques account for 227 out of 294 MITRE ATT\&CK mappings, covering 77.21\% of all technique mappings. The most frequent techniques are \texttt{T1499} Endpoint Denial of Service, \texttt{T1203} Exploitation for Client Execution, and \texttt{T1210} Exploitation of Remote Services. These results show that denial of service, client-side exploitation, and remote-service exploitation are recurring attack behaviors in CAV-related vulnerabilities. The CWE and MITRE ATT\&CK findings suggest that \gls{cwe} categories often lead to repeated exploitation and impact patterns in the transportation security domain.

\begin{table}[t]
\centering
\caption{Top 10 ATT\&CK Techniques in Our Dataset}
\label{tab:top10_mitre_attack_patterns}
\scriptsize
\resizebox{\columnwidth}{!}{
\begin{tabular}{llcc}
\hline
\textbf{Rank} & \textbf{Technique} & \textbf{Name} & \textbf{Count} \\
\hline
1  & T1499 & Endpoint Denial of Service & 48 \\
2  & T1203 & Exploitation for Client Execution & 42 \\
3  & T1210 & Exploitation of Remote Services & 34 \\
4  & T1200 & Hardware Additions & 19 \\
5  & T1557 & Adversary-in-the-Middle & 18 \\
6  & T1078 & Valid Accounts & 14 \\
7  & T1005 & Data from Local System & 14 \\
8  & T1068 & Exploitation for Privilege Escalation & 13 \\
9  & T1552 & Unsecured Credentials & 13 \\
10 & T1548 & Abuse Elevation Control Mechanism & 12 \\
\hline
\multicolumn{3}{l}{\textbf{Total top-10 technique occurrences}} & \textbf{227} \\
\multicolumn{3}{l}{\textbf{Coverage over all MITRE mappings}} & \textbf{227/294 = 77.21\%} \\
\hline
\end{tabular}
}
\end{table}

The co-occurrence results further show that \texttt{T1203--T1499} is the most frequent technique pair, appearing in 10 CVEs with support 0.0556. The next most frequent pairs are \texttt{T1210--T1499} with 6 co-occurrences and \texttt{T1068--T1548} with 5 co-occurrences. The association-rule results also show a bidirectional association between \texttt{T1548} Abuse Elevation Control Mechanism and \texttt{T1068} Exploitation for Privilege Escalation, with confidence values of 0.4167 and 0.3846. The most frequent MITRE ATT\&CK techniques are \texttt{T1499}, \texttt{T1203}, and \texttt{T1210}. The most common co-occurring technique pair is \texttt{T1203--T1499}, followed by \texttt{T1210--T1499} and \texttt{T1068--T1548}. 

\section{Discussion}
\label{sec:findings_discussion}

Our results provide key findings about open-weight LLM-based CVE-to-STIX generation for CAV vulnerabilities.

\textbf{Implication 1: Prompt guidance improves CVE-to-STIX generation.}
Contextless prompting provides a limited baseline because models receive minimal task instructions. Models can extract directly stated vulnerability objects, but they often miss implicit STIX objects, relationships, CWE mappings, and MITRE ATT\&CK mappings. STIX-guided prompting improves structural completeness through explicit instructions, while dynamic few-shot prompting achieves the highest overall performance in our results. These results indicate that reliable CVE-to-STIX generation requires both schema guidance and target-relevant examples.

\textbf{Implication 2: Object extraction performs better than relationship extraction.}
STIX object extraction reaches high performance under guided and example-based prompting, with multiple models reaching F1 scores around 0.94. Relationship extraction remains more complex because models must identify the correct source object, target object, and relationship type. Qwen3-Coder-30B achieves the highest relationship extraction F1 score of 0.6298 under dynamic few-shot prompting. This gap shows that CVE-to-STIX generation requires graph-level reasoning beyond entity recognition.

\textbf{Implication 3: CWE and MITRE ATT\&CK mappings require different reasoning.}
CWE mapping improves under dynamic few-shot prompting because similar examples help models connect vulnerability behavior with standardized weakness categories. MITRE ATT\&CK mapping remains more demanding because one CVE can correspond to multiple valid attack techniques. Match@1 improves for models such as Gemma4 and Qwen3-Coder-30B, but Match@All remains lower. These results suggest that future systems should combine LLM generation with external validation from CWE and MITRE ATT\&CK knowledge sources.

\textbf{Implication 4: Multi-agent generation provides task-specific gains.}
The Gemma4-based multi-agent configuration performs better for entity extraction, while the Codestral-22B-based configuration performs better for relationship extraction. However, multi-agent decomposition does not consistently outperform the strongest dynamic few-shot setting. This result indicates that role-based decomposition can help selected subtasks, but the benefit depends on the base model and agent role.

\textbf{Implication 5: CAV vulnerabilities contain recurring weakness and attack-behavior patterns.}
The ground-truth dataset includes frequent CWE categories that overlap with the 2025 CWE Top 25 list, including CWE-787, CWE-284, CWE-20, CWE-476, and CWE-863. The MITRE ATT\&CK analysis identifies T1499, T1203, and T1210 as the most frequent techniques, while T1203--T1499 is the most frequent co-occurring pair. These patterns show that CAV-related vulnerabilities often involve connected access-control, input-validation, exploitation, denial-of-service, and privilege-escalation behaviors. The dataset-level findings support the use of CAV-STIXGen for both LLM evaluation and transportation-domain threat analysis.

\section{Conclusion and Future Work}
\label{sec:conclusion}

We presented \textit{CAV-STIXGen}, a CAV-focused dataset for evaluating LLM-based CVE-to-STIX generation. The dataset maps CAV-related CVE descriptions to validated \gls{sdo}, \gls{sro}, \gls{cwe}, and MITRE ATT\&CK technique mappings. Using this dataset, we evaluated 11 open-weight LLMs ranging from 4B to 120B parameters across prompting strategies, temperature configurations, and multi-agent settings. The best single-model results reached 0.94 F1 for SDO extraction, 0.63 F1 for SRO extraction, and 0.99 F1 for CWE mapping, while complete ATT\&CK mapping remained more challenging because one CVE can involve multiple valid attack techniques. Our results show that prompt guidance improves CVE-to-STIX generation. Contextless prompting produces partial STIX representations, STIX-guided prompting improves structural completeness, and dynamic few-shot prompting achieves the strongest overall performance in our evaluation. The multi-agent setting provides task-specific gains, with Gemma-4-31B achieving 0.91 F1 for SDO extraction and Codestral-22B achieving 0.43 F1 for SRO extraction. Dataset-level analysis further identifies recurring weakness and attack-behavior patterns in the CAV domain, including overlap with the CWE Top 25 and repeated MITRE ATT\&CK co-occurrence patterns. 

Future work will extend \textit{CAV-STIXGen} with additional CAV-related vulnerabilities, richer STIX object types, and more detailed relationship annotations. Future studies can also evaluate retrieval-augmented prompting with external STIX, CWE, and MITRE ATT\&CK knowledge bases. Another direction is to improve relationship construction through graph validation, source-target consistency checking, and post-generation repair. Future multi-agent pipelines can assign specialized models to entity extraction, relationship construction, CWE mapping, MITRE ATT\&CK mapping, and validation. These extensions can improve the reliability of AI-assisted CVE-to-STIX generation for transportation-domain threat analysis.
 

\bibliographystyle{ACM-Reference-Format}
\bibliography{references}
\end{document}